\documentclass[conference]{IEEEtran}
\usepackage[colorlinks, citecolor=blue]{hyperref}
\usepackage{setspace, amsmath, amssymb, url, lscape, algorithmic, multirow, pslatex, listings, verbatim, alltt, amsfonts, wrapfig, boxedminipage, color, hyperref, bookmark}
\usepackage{cite}
\usepackage{graphicx}
\usepackage{textcomp}
\usepackage{xcolor}
\usepackage{caption}
\usepackage{subcaption}
\usepackage[vlined,linesnumbered,ruled,boxed]{algorithm2e}
\usepackage{balance} % For balanced columns on the last page

\newcommand{\paperfolder}{.}

\newcommand{\eg}{{\it e.g., }}
\newcommand{\etal}{{\it et~al. }}
\newcommand{\ie}{{\it i.e., }}

\newcommand{\comments}[1]{}

\def\BibTeX{{\rm B\kern-.05em{\sc i\kern-.025em b}\kern-.08em
    T\kern-.1667em\lower.7ex\hbox{E}\kern-.125emX}}
\begin{document}

\title{Robust Dynamic Resource Allocation via Probabilistic Task Pruning in Heterogeneous Computing Systems}

\author{%
 James Gentry, Chavit Denninnart, Mohsen Amini Salehi 
\\ High Performance Cloud Computing (HPCC) Laboratory,\\ School of Computing and Informatics, University of Louisiana at Lafayette, USA 
\\ \{cxd9974, amini\}@louisiana.edu 

}

\maketitle

\begin{abstract}
In heterogeneous distributed computing (HC) systems, diversity can exist in both computational resources and arriving tasks. In an inconsistently heterogeneous computing system, task types have different execution times on heterogeneous machines. A method is required to map arriving tasks to machines based on machine availability and performance, maximizing the number of tasks meeting deadlines (defined as \emph{robustness}). For tasks with hard deadlines (\eg those in live video streaming), tasks that miss their deadlines are dropped. The problem investigated in this research is maximizing the robustness of an oversubscribed HC system. A way to maximize this robustness is to prune (\ie defer or drop) tasks with low probability of meeting their deadlines to increase the probability of other tasks meeting their deadlines. In this paper, we first provide a mathematical model to estimate a task's probability of meeting its deadline in the presence of task dropping. We then investigate methods for engaging probabilistic dropping and we find thresholds for dropping and deferring. Next, we develop a pruning-aware mapping heuristic and extend it to engender fairness across various task types. We show the cost benefit of using probabilistic pruning in an HC system. Simulation results, harnessing a selection of mapping heuristics, show efficacy of the pruning mechanism in improving robustness (on average by $\simeq$25\%) and cost in an oversubscribed HC system by up to $\simeq$40\%.

%Second, we propose a method to set dropping and deferring thresholds to maximize the overall robustness of the HC system. Third, we compare two methods for deciding when to engage task-dropping, and examine the effect of accounting for history in the decision. Fourth, we propose a method to dynamically modify the dropping threshold for individual tasks. Fifth, we attempt to engender fairness amongst the task types by relaxing thresholds for task types that suffer dropping. Finally, we show the cost benefits of using probabilistic pruning in a cloud HC system. Simulation results, harnessing a selection of mapping heuristics, show the efficacy of our proposed pruning mechanism that can improve robustness of an oversubscribed HC system by up to XX\%. 
\end{abstract}


\section{Introduction}\label{sec:intro}
A Heterogeneous Computing (HC) system can be described by two types of heterogeneity: inconsistent and consistent~\cite{ali2000,li2018cost}. Inconsistent machine heterogeneity refers to differences in machine architecture (\eg CPU versus GPU versus FPGA~\cite{zahaf2017het, zhao2017fpga,hong2017gpu}). Consistent machine heterogeneity describes the differences among machines of a certain architecture (\eg different clock speeds). Compute services offered by cloud providers are a good example of an HC system. Amazon cloud~\cite{aws} offers inconsistent heterogeneity in form of various Virtual Machine (VM) types, such as CPU-Optimized, Memory-Optimized, Disk-Optimized, and Accelerated Computing (GPU and FPGA). Within each type, various VMs are offered with consistent performance scaling with price~\cite{aws}. Moreover, both consistent and inconsistent heterogeneity can exist in arriving tasks. For example, an HC system dedicated to processing live video streams is responsible for many categorically different types of tasks: changing video stream resolution, changing the compression standard, changing video bit-rate~\cite{li2018cost}. Each of these task types can be consistently heterogeneous within itself (\eg it takes longer to change resolution of 10 seconds of video, compared to 5).

Many HC systems (\eg~\cite{zong2017marcher,LONI}) present both consistent and inconsistent heterogeneity in machines used and task types processed~\cite{smith09}. These systems present cases where each task type can execute differently on each machine type, where machine type $A$ performs task type 1 faster than machine type $B$ does, but is slower than other machine types for task type 2. Specifically, compute intensive tasks run faster on (\ie matches better with) a GPU machine whereas tasks with memory and disk accesses bottlenecks (\eg in-memory databases~\cite{wang2014using,dos2015smart,malensek2016minerva}) runs faster on a CPU-based machine. 

All of this heterogeneity results in uncertainty for a given task's execution time, thus, inefficiency of resource allocation~\cite{ali2000}. Accordingly, a major challenge in HC systems is to assign tasks to machines to optimize performance goal of the system~\cite{ali2000}. We define \emph{robustness} as the degree to which a system can maintain performance in the face of uncertainty~\cite{Shestak08}. The overall \emph{goal} of this study is to maximize the robustness of an HC system.  

\begin{comment}
MOVED TO SYSTEM MODEL SECTION

\begin{figure}[htbp]
  \centering
  \includegraphics[width=0.4\textwidth]{\paperfolder/Figures/ResourceMapperArchitecture.pdf}
  \caption{\small{Arriving tasks enter a batch queue. Heterogeneous tasks are mapped to inconsistent heterogeneous machines in batches. In each mapping event, the pruner drops or defers tasks based on the tasks' probability of success.}  \label{fig:introQueue}}
\end{figure}
\end{comment}

Each task is considered to have a hard individual deadline, past which, no value remains in executing the task. Hence, tasks are dropped (\ie removed) from the system when their deadline passes~\cite{khemka2015utility,khemka14utility}. When the HC system is under load, such that it is impossible for all tasks to complete before their deadlines, the system is considered \emph{oversubscribed}. The performance metric based on which we measure robustness of an HC system is the number of tasks that meet their deadlines in the system. Therefore, the specific goal of this study is to maximize the number of tasks meeting their deadlines in the HC system (referred to as \emph{task success}) in the face of uncertain execution times in an oversubscribed system. A model of machine and task heterogeneity~\cite{alebrahim2017het} must be available to the resource allocation system, and the system must harness this awareness to overcome with the uncertainty of the HC system. 

When tasks have hard deadlines, time spent executing tasks that are ultimately dropped is wasted time. This wasted time cascades down the queue of tasks, delaying the execution of other tasks, and increasing the number of missed tasks in the future---decreasing system robustness. To mitigate this, tasks with a low probability of success should not be mapped, and if they are, they should be dropped before execution. If probabilistically pruning these unlikely-to-succeed tasks yields more tasks completing on-time in oversubscribed HC systems, how do we maximize the robustness gained thereby? %The more specific research questions can be stated as: \emph{(1) Below what level of probability should we defer tasks? (2) When should the resource allocation transition to a more aggressive mode and drop tasks?}

To address this question, in this research, we propose a pruning mechanism (as depicted in Figure~\ref{fig:introQueue}) that is composed of two methods, namely \emph{deferring} and \emph{dropping}. Task deferring deals with postponing assignment of unlikely-to-succeed tasks to a next mapping event with the hope that the tasks can be mapped to a machine that provides a higher chance of success for them. Alternatively, when the system is oversubscribed, the pruning mechanism transitions to a more aggressive mode and drops the tasks that are unlikely to succeed. Before determining deferring and dropping details, we need to model the impact of task dropping on the probability of success for the tasks behind the dropped task. Then, we determine the appropriate probability for dropping and deferring. We propose a method to dynamically determine when the resource allocation system should transition to a more aggressive mode and engage in task dropping. We compare and analyze robustness obtained from deploying our proposed pruning mechanism against an HC system that either does not perform pruning or has a basic pruning implemented. 

Maximizing robustness of HC systems in terms number of tasks meeting their deadlines can potentially cause bias towards executing certain task types and affects fairness of the system. As such, we develop a mapping method to maintain fairness while maximizing robustness.% and compare it against other fairness-oblivious methods. 

Our hypothesis is that the proposed pruning mechanism not only improves robustness of an HC system, but can impact the incurred cost of using resources. This is particularly important for users who deploy heterogeneous cloud VMs~\cite{salehi10}.
As such, we investigate the impact of the proposed probabilistic pruning mechanism on the incurred cost of using heterogeneous cloud VMs and compare it against common mapping methods. 

In summary, the contributions of this paper are as follows:
\begin{itemize}
	\item Mathematically modeling the impact of task dropping on the success probability of other tasks behind the dropped one.
	
	\item Developing a method to determine probablistic dropping and deferring thresholds in the pruning mechanism.
	%\item Introduce and study a dynamic, per-task pruning threshold.
	\item Proposing a method to dynamically engage task dropping in response to oversubscription in an HC system.
	\item Developing a pruning-aware and a fairness-aware mapping heuristic for an HC system.
	\item Analyzing the cost benefit of the pruning mechanism.
\end{itemize}

Simulation results approve our hypotheses and show that the pruning mechanism can enhance robustness and the incurred cost. Importantly, the mechanism is more impactful under higher oversubscription levels. This paper is organized as follows. Section \ref{sec:related} situates this work in relation to existing literature. Section \ref{sec:background} establishes the problem and describes our system model. Then, Section \ref{sec:solution} presents our proposed solution. Baseline heuristics are described in Section \ref{sec:experiment}, along with the constraints and parameters of the experiment. Section \ref{sec:evltn} presents and analyzes the simulation results. Finally, Section \ref{sec:conclsn} concludes the paper and offers direction for future works.

\section{Related Works}\label{sec:related}
Mapping tasks in HC systems have been shown to be an NP-complete problem~\cite{coffman76,Ibarra77}. As such, there are multiple prior efforts that achieve sub-optimal solutions and they are either influence or being similar to our work. Here are some notable mentions.

%used in infocom, modded, important
To model task execution times, Shestak \etal~\cite{Shestak08}, instead of using a scalar value, lay the groundwork for the use of probability mass functions (aka PMF). The method for convolution of execution times to form completion times for a queue of tasks is established. Our work builds upon their use of PMFs and robustness measurement, while also adding the conditions of probabilistically drop executing tasks and pending tasks.
Khemka \etal~\cite{khemka2015utility} investigate resource allocation in oversubscribed heterogeneous systems. They test task utility functions based on priority, utility class, and urgency. They use a matrix with deterministic execution times, whereas we model the times probabilistically. Also, unlike our approach of probabilistically determining if a task should be dropped, their task dropping occurs only after a task's utility goes below a static threshold. In~\cite{salehi2016stochastic}, Salehi \etal model the stochastic nature of the heterogeneous task types on heterogeneous machine types using a matrix of probability mass functions (PMFs) to improve robustness of dynamic resource allocation.  A mathematical model for calculating the completion time of stochastically modeled tasks in the presence of task dropping is provided. However, Salehi \etal only consider dropping tasks after their deadlines have passed. %Our work builds upon the matrix of PMFs, and the completion time calculations to enable probabilistic task dropping. 

%% original below
%In~\cite{malawski2015algorithms}, Malawski \etal evaluate dynamic scheduling of deadline-and-cost constrained tasks in IaaS clouds. The work is focused on provisioning homogenous VMs, whereas we focus on static heterogeneous machines. The goal of the study is maximizing the high priority workflows completed, within a deadline, while remaining under budget. Algorithms are introduced that are aware of the workflow, leading to higher performance, and they drop workflows that would result in a loss of high priority tasks completing. We use a similar tactic of dropping tasks to maximize the robustness of our system, however our metrics for success are different, and do not consider priority. 

%from Thesis, modded, commented out to save space, bring back 2018-oct-01

%used in infocom and ucc, modded
Delimitrou and Kozyrakis~\cite{paragon13} propose Paragon which is an immediate (\ie not batch) dynamic scheduling system for heterogeneous data centers. They use singular value decomposition of historical data to classify incoming tasks based on their heterogeneity. The classifications are used in a greedy algorithm to select a list of candidate resources based on interference, and then from that, the best fit based on heterogeneity~\cite{delimitrou2014quality}. 
Unlike our work that considers probabilistic execution times for decision making, their mapping heuristics operates based on scalar execution times. The performance metrics are also different, as their tasks do not have deadline to consider, Paragon is only concerned system throughput.
%%%%%%%original from infocom
%Paragon~\cite{paragon13} is an immediate mode scheduling system for large-scale heterogeneous datacenters. It uses singular value decomposition to classify incoming tasks on their heterogeneity, as well as their intereference level for co-scheduled tasks. This is accomplished via the matching of information from historical data, with small sample runs of the task. These classifications are used in a greedy algorith to select a list of candidate resources first based on interference, and from that the best fit based on heterogeneity~\cite{delimitrou2014quality}. This work is different than ours in that the mapping heuristics are immediate mode using scalar --as opposed to probabilistic-- execution times to make decisions. The performance metrics are also different, as Paragon is concerned with speedup, as there are no deadlines to miss.

In~\cite{liperformanceanalysis}, Li \etal introduce the affinity (\ie match) of heterogeneous cloud VMs to change coding of video streams. They observed that depending on their content types, video files have different performances on heterogeneous VM types. Particularly, they notice that slow-motion video contents gain from compute intensive VMs, such as GPUs, whereas fast-motion videos do not gain much from such VMs.
They concluded that categorizing videos based on their content types and deploying an inconsistently heterogeneous set of cloud VMs can reduce the incurred cost of using cloud without compromising quality. In another work~\cite{li2018cost}, Li \etal dynamically composes an inconsistently HC system to process a heterogeneous set of video streaming tasks. However, they do not consider the case of task dropping.% because they focus on Video On Demand where tasks continue until completion. 

Malawski \etal~\cite{malawski2015algorithms} evaluate dynamic mapping of deadline- and cost-constrained tasks in cloud. They support dropping workflows that would result in a loss of high priority tasks completion, however, their metrics to quantify and evaluate each task's worthiness are different. Unlike our work, they focus on homogeneous cloud VMs. %, and do not have a priority consideration. 
Tetrisched~\cite{tumanov2016tetrisched} is a mapping method for consistent HC systems used for YARN and MapReduce. It operates based on mixed integer linear programming and considers task execution time on different machines types. Our system uses a similar set of information to for mapping, however, it also leverages task deferring to find a better match for tasks and considers task dropping to alleviate oversubscription and improve robustness. %Mage~\cite{romero2018mage} is a two-tier task mapper based on online machine learning techniques. It maps a task to a group of machines and then to a specific machine within the group. Our work differs from Mage because we consider inconsistent heterogeneity along with the pruning mechanism.

\section{System Model}\label{sec:background}
\begin{figure}[htbp]
  \centering
  \includegraphics[width=0.4\textwidth]{\paperfolder/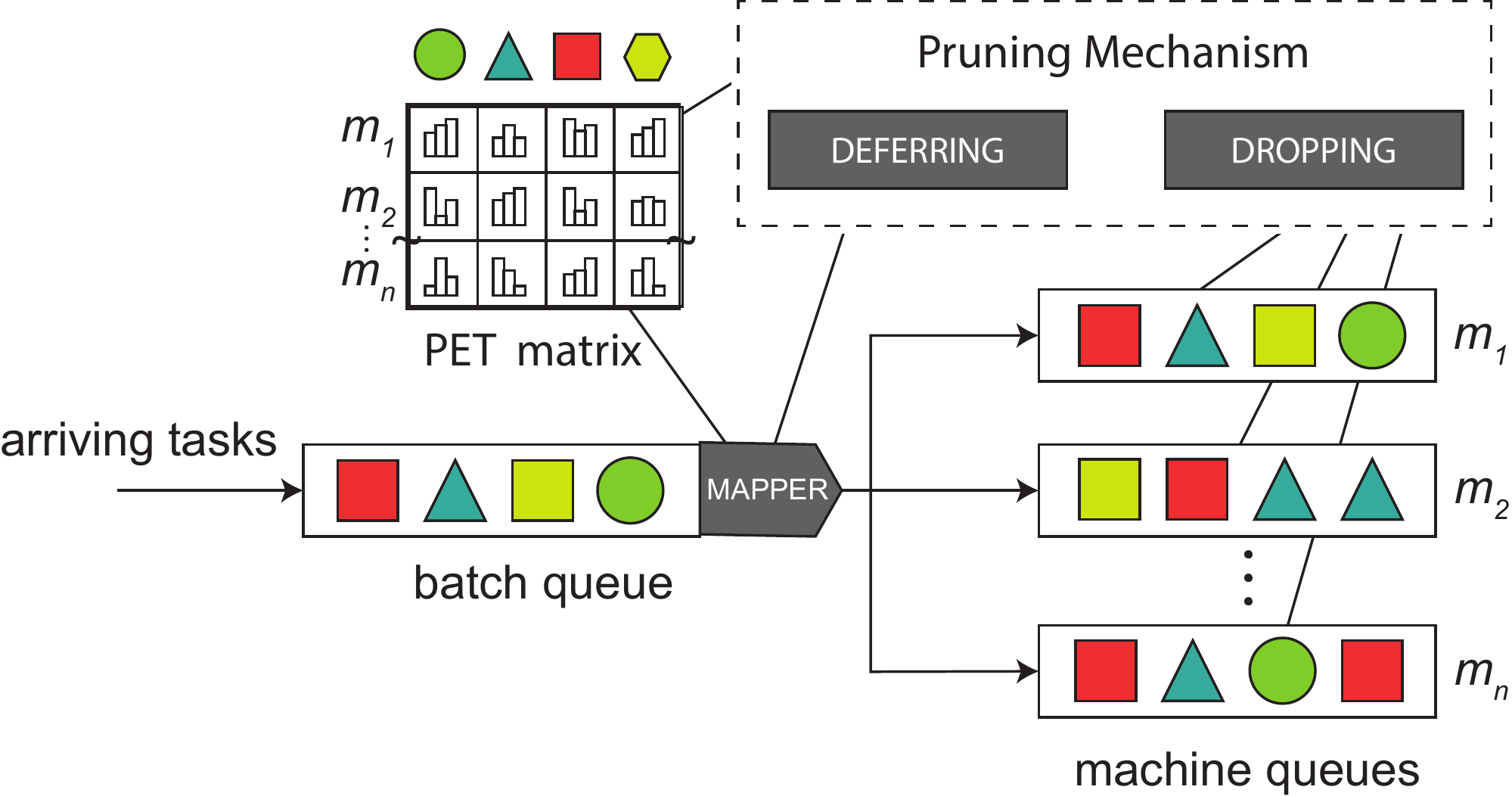}
  \caption{\small{Pruning mechanism. Heterogeneous tasks are mapped to heterogeneous machines in batches. In each mapping, the pruner drops or defers tasks based on their probability of success.}\label{fig:introQueue}}
\end{figure}

The motivation for this research comes from an HC system used for processing live video streaming services~\cite{li2016vlsc} (\eg YouTube Live and Twitch.tv~\cite{aparicio2015transcoding}). In these services, video content is initially captured in a certain format and then processed (aka transcoded) to support diverse viewers' display devices~\cite{li2016high}. As there is no value in executing live video streaming tasks that have missed their individual deadlines, they are dropped from the HC system. It has been shown that, in such a system deploying an inconsistently HC system helps processing inconsistently heterogeneous task types (\eg tasks to change resolution and tasks to change compression standard) and ensuring an uninterrupted streaming experience~\cite{liperformanceanalysis}. Figure~\ref{fig:introQueue} shows an overview of the system. Tasks are queued upon arrival and are mapped to available heterogeneous machines ($m_1, m_2,...,m_n$) in batches. 

To capture the stochastic nature in execution time of each task type (\eg those arising from data-size differences in tasks), we use Probability Mass Functions (PMF). In an inconsistently HC system, the execution time PMF of different task types on different machine types are maintained in a matrix called a \emph{Probabilistic Execution Time} (PET)~\cite{ali2000,salehi2016stochastic}. As we consider the HC systems are deployed to offer a specific service (\eg video streaming), the type of arriving task requests are limited and known. As such, the PET matrix has a limited and constant size. In practice, the PMFs of the PET matrix can be built from historic execution time information of each task type on each machine type and modeling them via a histogram in an offline manner~\cite{wasserman03}. Thus, we assume that such a PET matrix is available in our HC system. %\hl{This modeling can be done offline, and periodic on-line profiling can keep them consistent with any changes in the system}~\cite{somthingmaybe}. 

%\hl{Our system dynamically allocates available resources to incoming tasks.}
In our system, as seen in Figure~\ref{fig:introQueue}, heterogeneous tasks dynamically arrive into a batch queue of unmapped tasks with no prior knowledge of the timing or order. The intensity of tasks arriving to the HC system (\ie oversubscription level) also varies. To limit the compound uncertainty and maintain accuracy of mapping decisions, machines use limited-size local queues to process their assigned tasks in a first-come-first-serve (FCFS) manner. A mapping event occurs upon arrival of a new task or when a task gets completed. Before the mapping event, tasks that have missed their deadlines are dropped (removed) from the system. Then, the mapping event attempts to map tasks from the batch queue. This happens until either the machine queues are full, or there are no more unmapped tasks. We assume that once a task is mapped to a machine, its data is transferred to that machine and it cannot be remapped due to data transfer overhead. It is assumed that each task is independent and executes in isolation on a machine, with no preemption and no multitasking~\cite{dogan04,cao13}. 

To map tasks to machines, the mapper creates a temporary (virtual) queue of machine-task mappings and calculates the completion time distribution of each unmapped task on heterogeneous machines, as explained in the next section. 

\section{Calculating Task Completion Time in the Presence of Task Dropping}\label{sec:convolution}

Upon dropping a task in a given machine queue, the completion time PMF of those tasks behind the dropped tasks is improved. Intuitively, dropping a task, whose deadline has passed or has a low chance of success, enables the tasks behind it to begin execution sooner, thus, increasing their probability of success and subsequently, overall robustness of the HC system. Each task in queue compounds the uncertainty in the completion time of the tasks behind it in the queue. Dropping a task excludes its PET from the convolution process, reducing the compound uncertainty as well.  

The pruning mechanism we propose in this research should be able to calculate the impact of dropping a task on the probability of success (\ie robustness) of tasks behind the dropped tasks. In this section, we provide the mathematical model to calculate the completion time and probability of meeting deadline of a task located behind a dropped task.

%In the literature, the expected execution times of different task types on different machine types are maintained in an Expected Time to Compute (ETC) matrix~\cite{Maheswaran97}. These ETC matrices generally contain the mean execution time, and do not account for consistent task heterogeneity (\eg those arising from data-size differences across different tasks, environmental factors such as neighboring loads, task switching overhead, etc). The stochastic nature of the execution time of tasks that arises from these factors is modeled via Probability Mass Functions (PMF) called a \emph{Probabilistic Execution Time} (PET)~\cite{salehi2016stochastic}. In an inconsistently HC system, a PET matrix is maintained to describe probabilistic execution time of each task type on each machine type. The PET matrix is used by the task mapper to optimally map the tasks to machine. In practice, these PMFs can be obtained from historic execution time information of task types on each machine in an HC system and modeling them via a histogram~\cite{wasserman03}. The PET matrix is assumed to be available in our HC system. 

Recall that each entry $(i,j)$ of PET matrix is a PMF represents the \emph{execution time} of task $i$'s task type on a machine type $j$. In fact, $PET(i,j)$ is a set of impulses, denoted $E_{ij}$, where $e_{ij}(t)$ represents execution time probability of a single impulse at time $t$. Similarly, completion time PMF of task $i$ on machine $j$, denoted $PCT(i,j)$, is a set of impulses, denoted $C_{ij}$, where $c_{ij}(t)$ is an impulse represents the completion time of task $i$ on machine $j$ at time $t$. 

Let $i$ a task with deadline $\delta_i$ arrives at time $\alpha$ and is given a start time on idle machine $j$. In this case, the impulses in $PET(i,j)$ are shifted by $\alpha$ to form its $PCT(i,j)$~\cite{salehi2016stochastic}. %Let $c_{ij}(t)$ an impulse in the completion time PMF of task $i$ on machine $j$ at time $t$. 
Then, the robustness of task $i$ on machine $j$ is the probability of completing $i$ before its deadline, denoted $p_{ij}(\delta_i)$, and is calculated based on Equation~\ref{eq:0}. 

\begin{equation} \label{eq:0}
    p_{ij}(\delta_i) = \sum_{t=\alpha}^{t\leq \delta_i} c_{ij}(t)
\end{equation}
\vspace{2pt}

\begin{figure}[htbp]
\vspace{-2pt}
  \centering
  \includegraphics[width=0.5\textwidth]{\paperfolder/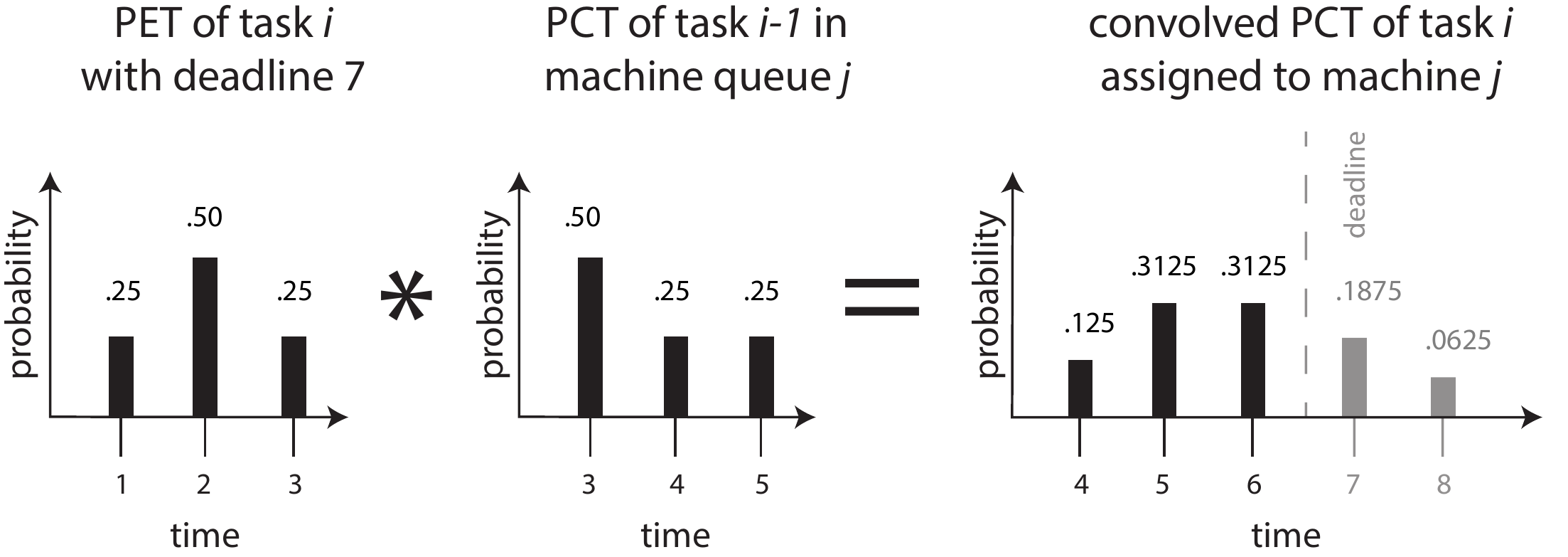}
  \caption{\small{Probabilistic Execution Time (PET) of arriving task $i$ is convolved with the Probabilistic Completion Time (PCT) of the last task on machine $j$ to form $PCT(i,j)$.} \label{fig:conv} }

\end{figure}

In case machine $j$ is not idle (\ie it has executing or pending tasks) and task $i$ arrives, the PCT of the last task in machine $j$ (\ie $PCT(i-1,j)$) and $PET(i,j)$ are convolved to form $PCT(i,j)$. This new PMF accounts for execution times of all tasks ahead of task $i$ in the machine queue $j$. For example, in Figure~\ref{fig:conv}, an arriving task $i$ with $\delta_i=7$ is assigned to machine $j$. Then, $PET(i,j)$ is convolved with the PCT of the last task on machine queue $j$ to form $PCT(i,j)$. 

%In order to calculate the completion time distribution (CTD) to evaluate possible task-machine mapping decisions,  For any PCT generation, the presence of past-deadline task dropping needs to be taken into account, and if present, the type of task dropping must also be accounted for.

The completion time impulses are generated differently based on the way task dropping is permitted in a system. Three scenarios are possible: \textbf{(A)} Task dropping is not permitted; \textbf{(B)} Only pending tasks can be dropped; and \textbf{(C)} Any task, including the executing one, can be dropped. We note that the initial idea of calculating these completion time PMFs were proposed in~\cite{salehi2016stochastic}. However, in the following, we mathematically model and provide the closed form solution for calculating completion time PMFs. Considering the space limit, interested readers can refer to~\cite{salehi2016stochastic} for further explanations.

\textbf{(A)} Task dropping is not permitted, \ie when all mapped tasks must execute to completion, Equation~\ref{eq:1} is used to calculate the impulses, denoted $c_{ij}^{NoDrop}(t)$, of $C_{ij}$ from the convolution of $PET(i,j)$ and $PCT(i-1,j)$.

\begin{equation} \label{eq:1}
    c_{ij}^{NoDrop}(t)= \sum_{k=1}^{k<t}[e_{ij}(k) \cdotp c_{(i-1)j}^{NoDrop}(t-k)]
\end{equation}
\vspace{3pt}

%$c_{ij}(t)$ is an impulse in the completion time convolution of two probability distributions. 
%It is the probability that task i will complete at time x. It is obtained by convolving task i's execution time distribution with the completion time distribution of task i-1. $C_{i-1}(x)$ is the same probability, but for the task before task i in queue, and so on. The first task has its execution time distribution convolved with one, as it is guaranteed to start then. 

%For the next few equations, the $\delta$ is the deadline of task i-1.
\textbf{(B)} Only pending tasks can be dropped. In this case, the impulses in $PCT(i-1,j)$ that occur after the deadline of task $i$ are not considered in calculating $PCT(i,j)$, as that would indicate task $i$ is dropped due to its deadline passing. Therefore, the formulation changes to reflect the impact of truncated $PCT(i-1,j)$ in the convolution process. Owing to the complexity of calculating $PCT(i,j)$, in this circumstance, we develop a helper function, denoted $f(t,k)$, as shown in Equation~\ref{eq:helper}, that helps Equation~\ref{eq:pending} to discard impulses from  $PCT(i-1,j) \geq \delta_i$. To calculate impulse $c_{ij}(t)$, note that if $t < \delta_i$, then $t-k <\delta_i$. In this case, Equations~\ref{eq:pending} and~\ref{eq:helper} operate the same as Equation~\ref{eq:1}. However, for cases where $t \geq \delta_i$, we use the helper Equation~\ref{eq:helper} to generate an impulse by discarding impulses of  $PCT(i-1,j)\geq \delta_i$. Later, in Equation~\ref{eq:pending}, we add impulses in $i-1$ that occur after $\delta_i$ to account for when task $i-1$ completes at or after $\delta_i$.

\begin{equation} \label{eq:helper}
    f(t,k)= 
\begin{cases}
    0, & \forall (t - k)\geq \delta_i\\
    \\
    e_{ij}(k) \cdotp c_{(i-1)j}^{pend}(t - k),& \forall (t - k)< \delta_i
\end{cases}
\end{equation}
%\vspace{14pt}

\begin{equation} \label{eq:pending}
    c_{ij}^{pend}(t)= 
\begin{cases}
    \sum\limits_{k=1}^{k<t}f(t,k)+c_{(i-1)j}^{pend}(t), & \forall t\geq \delta_i\\
    \\
    \sum\limits_{k=1}^{k<t}f(t,k), & \forall t< \delta_i
\end{cases}
\end{equation}
\vspace{3pt}

\textbf{(C)} All tasks (including executing one) can be dropped. In fact, in this case, the completion time impulses are obtained similar to Equation~\ref{eq:pending}. However, the special case happens when $t=\delta_i$ because at this time, if task $i$ has not completed, it is dropped. For the purposes of calculating $PCT(i,j)$ using Equation~\ref{eq:5}, $PCT(i-1,j)$ is guaranteed to be complete by its deadline. Therefore, as Equation~\ref{eq:5} shows, all the impulses after $\delta_i$ are aggregated into the impulse at $t=\delta_i$. We should note that, the discarded impulses, \ie those of task $i-1$ that occur at or after $\delta_i$, must be added to $C_{ij}$, to indicate the probabilities that task $i-1$ completes after task $i$'s deadline.

\begin{equation} \label{eq:5}
    c_{ij}^{evict}(t)= 
\begin{cases}
\sum\limits_{k=t}^{k<\infty}c_{ij}^{pend}(k) + c_{(i-1)j}^{evict}(t), &  t= \delta_i\\
\\
   c_{ij}^{pend}(t), & \forall t> \delta_i\\
   \\
   \sum\limits_{k=1}^{k<t}f(t,k), & \forall t< \delta_i
\end{cases}
\end{equation}
\vspace{3pt}

We note that, calculating completion time poses a not insignificant overhead. However, the overhead can be mitigated by pre-processing and memoizing portions of the convolution and finalizing it just in time at mapping time. It is also possible to approximate PMFs by aggregating impulses.

\section{Maximizing Robustness via Pruning Mechanism}\label{sec:solution}
\subsection{Overview}
In the beginning of the mapping event, if the system is identified as oversubscribed, the pruning mechanism (aka pruner) examines machine queues. Beginning at the executing task (queue head), for each task in a queue, the success probability (robustness) is calculated. Tasks whose robustness values are less than or equal to the dropping threshold are removed from the system. Then, the mapping method determines the best mapping for tasks in the batch queue. Prior to assigning the tasks to machines, the tasks with low chance of success are deferred (\ie not assigned to machines) and returned to the batch queue to be considered during the next mapping events. This is in an effort to increase robustness of the system by waiting for a machine with better match to become available for processing the deferred task. To design the pruner for an HC system, three sets of questions regarding deferring and dropping operations are posed that need to be addressed. 

\emph{First} set of questions surround the probability thresholds at which tasks are dropped or deferred. How to identify these thresholds and the relation between them. A related question arises is, should a system-level probability threshold be applied for task dropping? Or, should there be individual considerations based on the characteristics of each task? If so, what characteristics should be considered, and how should they be used in the final determination?
%The system uses the number of tasks that have missed their deadlines since the previous mapping event as a toggle for the dropping mechanism. The question is should the same value be used to start and stop dropping, or should two values be used (\ie as in a Schmitt Trigger Circuit)~\ref{schmitt1938} Should the system give some weight to previous mapping event's deadline misses as well, or should each decision be made independently?
\emph{Second}, there is the matter of when to begin task dropping, and when to cease. That is, how to dynamically determine the system is oversubscribed and transition the pruner to a more aggressive mode to drop unlikely-to-succeed tasks such that the overall system robustness is improved. 
Pruning can potentially lead to unfair scheduling across tasks types---constantly pruning compute-intensive and urgent task types in favor of other tasks to maximize the overall robustness. Hence, the \emph{third} question is how the unfairness across task types can be prevented? Should the system prioritize task types that have been pruned? If so, how much of a concession should be made?

\subsection{Determining Dropping and Deferring Thresholds}\label{subsec:dropdefer}
\subsubsection{Dynamic Per-Task Dropping Threshold}
At its simplest, the task dropper can apply uniform dropping threshold for all tasks in a machine queue. However, a deeper analysis tells us that not all tasks have the same effects on the probability of on-time completion for the tasks behind them in queue. This can be taken into account to make the best decision about which tasks should stay and which are dropped. 

%As we mentioned earlier, the completion time PMF of a given task $i$ is used for calculating the task robustness. 
In addition to determining task robustness, other features of completion time PMF can be valuable in making decisions about probabilistic task dropping. We identify two task-level characteristics that further influence the robustness of tasks located behind a given task $i$: (A) the position of task $i$ in machine queue, and (B) the shape (\ie skewness) of completion time PMF of task $i$. 

In fact, the closer a task is to execution, the more tasks are affected by its completion time. For instance, with a machine queue size of six, an executing task affects the completion time of five tasks queued behind it, where the execution time of a task at the end of the queue affects no tasks. Therefore, the system should apply a higher dropping threshold for tasks close to queue head. 

Skewness is defined as the measure of asymmetry in a probability distribution and is calculated based on Equation~\ref{eq:skew}~\cite{bayes2007bayesian}. In this equation, $N$ is the sample size of a given PMF, $Y_i$ is an observation, and $\bar{Y}$ is the mean of observations. A negative skewness value means the tail is on the left side of a distribution whereas a positive value means that the tale is on the right side. Generally, $|S| \geq 1$ is considered highly skewed, thus, we define $s$ as bounded skewness and we have $-1 \leq s \leq 1$. 
\vspace{-2pt}
\begin{equation}\label{eq:skew}
S = \frac{\sqrt{N(N-1)}}{N-2} \times \frac{\sum_{i=1}^{n} {(Y_i-\bar{Y})^3}/N}{\sigma^3} 
\end{equation}
%\vspace{1pt}

\begin{figure} [htpb]
  \centering
  \begin{subfigure}[htpb]{0.47\textwidth}
  \centering
    \includegraphics[width=\textwidth]{\paperfolder/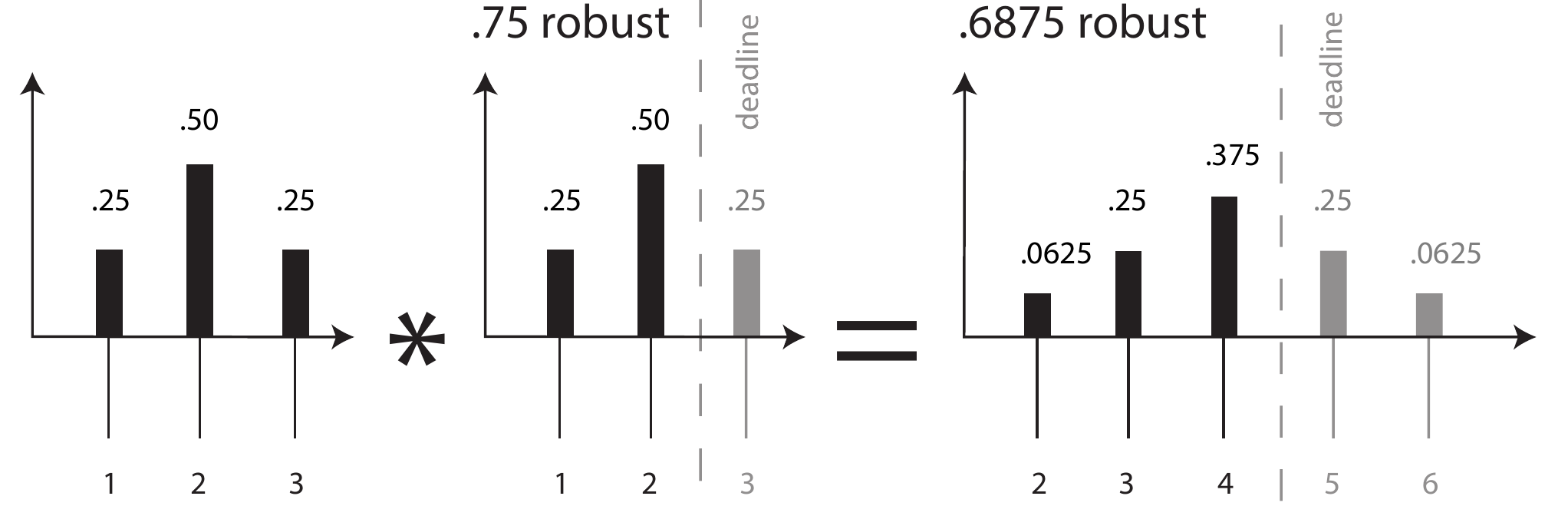}
    \caption{No Skew}
    \vspace{10px}
    \label{fig:noskew}
  \end{subfigure}

  \begin{subfigure}[htpb]{0.47\textwidth}
  \centering
    \includegraphics[width=\textwidth]{\paperfolder/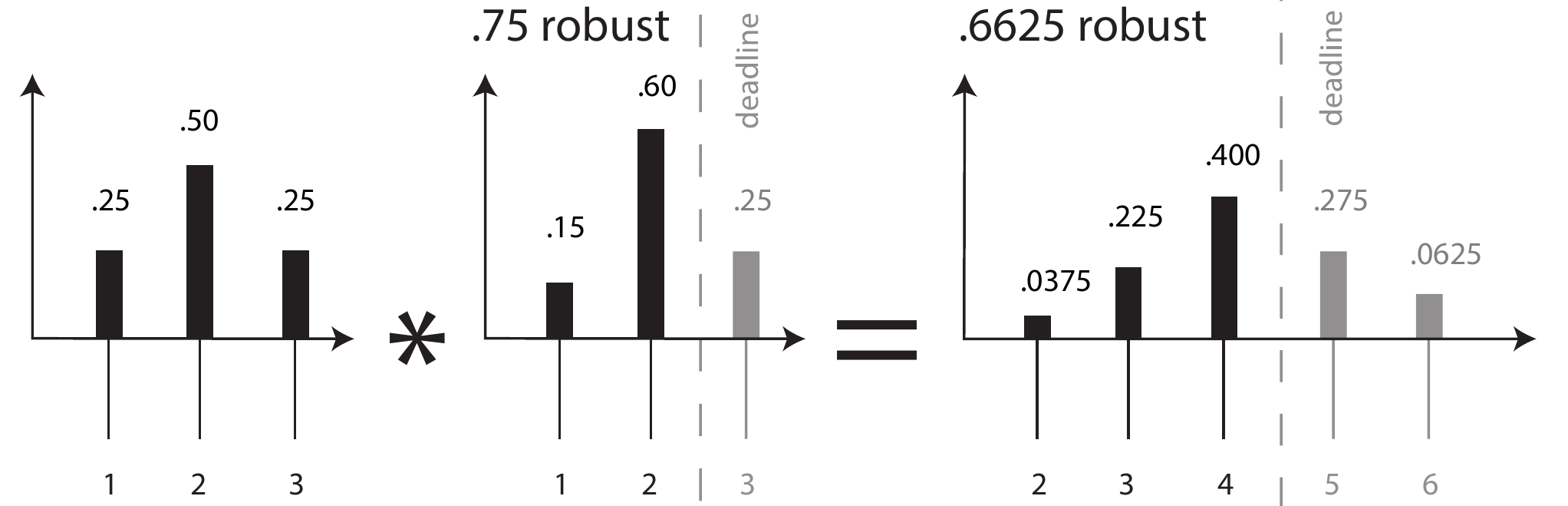}   
    \caption{Left Skew}
    \vspace{10px}
    \label{fig:leftskew}
  \end{subfigure}
  
  \begin{subfigure}[htpb]{0.47\textwidth}
  \centering
    \includegraphics[width=\textwidth]{\paperfolder/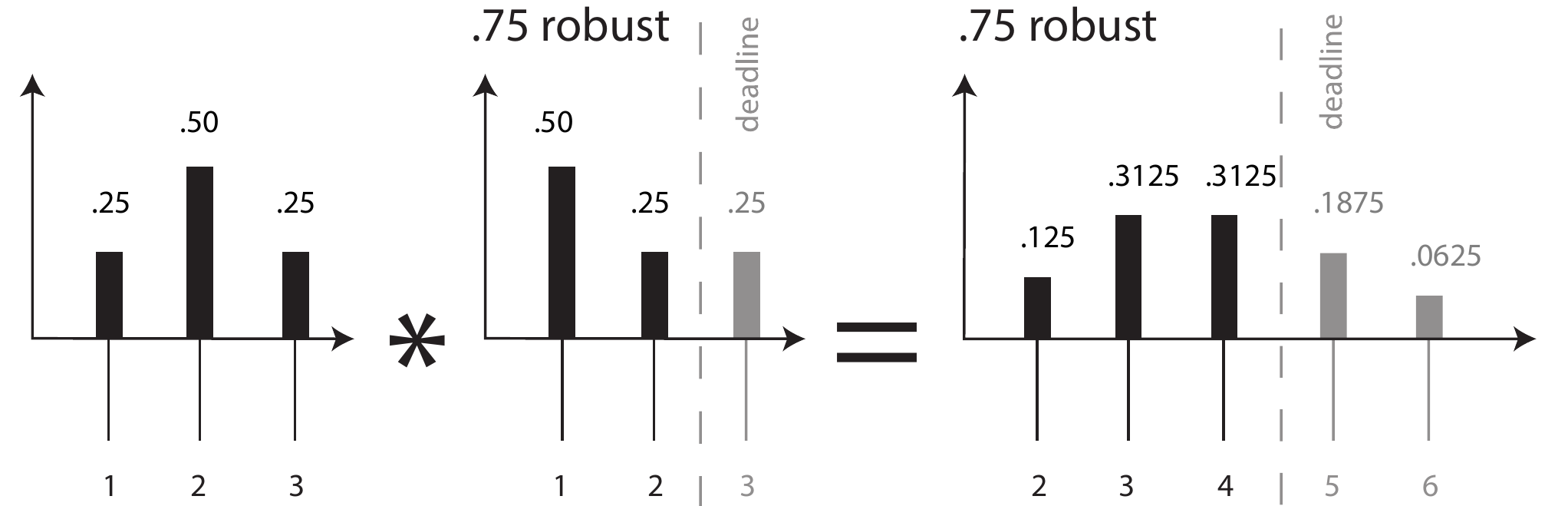}
    \caption{Right Skew}
    \vspace{5px}
    \label{fig:rightskew}
  \end{subfigure}
  
  \caption{\small{Demonstration of effect of task $i$'s skewness on completion time PMF of task $i+1$ (right-most PMFs) with a deadline 5 ($\delta_{i+1}=5$). The left-most PMFs show execution time PMF of task $i+1$ and the middle ones show completion time PMF of task $i$ ($\delta_i=3$).} \label{fig:skewsplain}}
\end{figure}

%The sign of skewness value denotes the direction of the skew. 
A negatively skewed PMF has the majority of probability occurring on the right side of PMF. Alternatively, because the bulk of a probability is biased to the left side of a PMF, a positive skew implies that a task is completed sooner than later. The middle PMFs in Figure~\ref{fig:skewsplain} each represents a completion time with a robustness of 0.75, however, they show different types of skewness. Using this information, we can see that two tasks with the same robustness can have different impacts on the robustness of tasks behind them in queue. Tasks that are more likely to complete sooner (\ie positive skewness) propagate that positive impact to tasks behind them in queue. The opposite is true for negatively skewed tasks. Reasonably, we can favor tasks with positive skewness in task dropping. Figure~\ref{fig:skewsplain} shows the effects of different types of skews on the completion times of tasks behind them in queue. Subfigure~\ref{fig:leftskew} shows the negative effects of negative skew whereas Subfigure~\ref{fig:rightskew} shows the positive effect of positive skew on the robustness of the next task in the queue. 
\vspace{-7pt}
\vspace{-3pt}

Using the skewness and queue position, the system can adjust a base dropping threshold dynamically, for each task in a machine queue. The adjusted dropping threshold for a given task $i$, denoted $\phi_i$, is calculated based on Equation~\ref{eq:location}. To favor tasks with positively skewed completion time PMF, we negate the skewness ($s_i$). To account for position of task $i$ in machine queue, denoted $\kappa_i$, we divide the negated skewness by the position. Addition of 1 is just to avoid division by zero and $\rho$ is a parameter to scale the adjusted dropping threshold. Ideally, this will allow more tasks to complete before their deadline, leading to a higher robustness in an HC system.
%\vspace{-1pt}
\begin{equation}\label{eq:location}
\phi_i=\frac{-s_i \cdotp \rho}{\kappa_i+1}
\end{equation}
%\vspace{1pt}

This dynamic adjustment of the probability is done only in the dropping stage of the pruner. When it comes to deferring tasks, the task position is always the same (\ie the tail of the queue), and it is too early to consider the shape of the tasks PMF, as there are, as yet, no tasks behind it in queue.

\subsubsection{Inferring Deferring Threshold from Dropping Threshold}
At its simplest, the pruner can use a single threshold to decide whether to defer mapping a task, as well as to decide whether to drop a mapped task. However, considering deferring threshold less than or equal to the dropping threshold, causes mapping a task with a robustness lower than the dropping threshold. Nonetheless, unless a task is dropped ahead of that mapped task, such a task is going to be dropped during the next engagement of the task dropper and leads to a poor performance. Therefore, considering deferring threshold \emph{higher} than the dropping threshold benefits the overall robustness of the system. This is because the pruner waits for more robust mapping decisions and considers a gap between the requirements to map a task (\ie deferring threshold), and the threshold required to dropping a task. In Section~\ref{subsec:thresh}, we explore the appropriate gap between deferring and dropping thresholds so that the robustness of HC system is maximized.

\subsection{Aggressive Pruning by Engaging Task Dropping}
To maximize robustness of the system, the aggression of the pruning mechanism has to be dynamically adjusted in reaction to the level of oversubscription in the HC system. 
The pruning mechanism considers the number of tasks missed their deadlines since the past mapping event as an indicator of the oversubscription level in the system. We use the identified oversubscription level as a \emph{toggle} that transitions the pruner to task dropping mode. 
However, in this case, the pruner can potentially toggle to dropping mode as a result of an acute spike in task-arrival, and not a sustained oversubscription state. %To avoid this, the Pruner's toggle does not react to temporary spikes in deadline misses, but only to sustained oversubscription by using a weighted average of deadline misses over the past mapping events. Thereby, the Pruner is slower to engage, and slower to disengage.

To judge the oversubscription state in the system, the pruner operates based on moving weighted average number of tasks that missed their deadlines during the past mapping events. Let $d_\tau$ the oversubscription level of the HC system at mapping event $\tau$; and \(\mu_{\tau}\) the number of tasks missing their deadline since the past mapping event. Parameter $\lambda$ is tunable and is determined based on the relative weight assigned to the past events. The oversubscription level is the calculated based on Equation~\ref{eq:oversub}. In the experiment section, we analyze the impact of $lambda$ and determine an appropriate value for it.

\begin{equation}\label{eq:oversub}
d_\tau=\mu_\tau \cdotp \lambda + d_{\tau-1} \cdotp (1-\lambda)
\end{equation}

Another potential concern is minor fluctuations about the toggle switching the dropping off and then back on. We employ a Schmitt Trigger~\cite{kader2012advancement} to prevent minor fluctuations around dropping toggle. We set separate on and off values for the dropping toggle. Based on our initial experiments, we determined the Schmitt Trigger to have 20\% separation between the on and off values. For instance, if oversubscription level two or higher signals starting dropping, oversubscription value 1.6 or lower signals stopping it.

\subsection{Proposed Mapping Heuristics}\label{subsec:heuristics}
In this part, we develop two mapping heuristics based on the theory explored in this study. The first heuristic, PAM, leverages pruning mechanism to maximize robustness. However, the second mapping heuristic, in addition to maximizing robustness, aims at achieving fairness across task types. 

The batch heuristics are two-phase processes, a first phase finds the best machine for each task, by virtue of a per-heuristic objective. In the second phase, from task-machine pairs obtained in the first phase, each heuristic chooses the best task-machine pair for each available machine queue slot. After all slots are filled, virtual mappings are assigned to the machine queues and the mapping method is complete.

\subsubsection{Pruning Aware Mapper (PAM)}
This heuristic uses the PET matrix to calculate task robustness and then operates based on the pruning mechanism. Before making any mapping decision, PAM analyzes the oversubscription level and performs task dropping on machine queues, if necessary. 

In the first phase, for each unmapped task, PAM finds the machine offers the highest robustness. Then, tasks that do not meet the deferring threshold are pruned. The second phase finds the task-machine pair with the lowest completion time and maps it to that machine's virtual queue. Ties are broken by choosing task with the shortest expected execution time. 
%The process repeats until either all tasks in the temporary batch queue are mapped or dropped, or until the virtual machine queues are full. 

\subsubsection{Fair Pruning Mapper (PAMF)}
Probabilistic task pruning potentially favors task types with shorter execution times, resulting in unfairness. This is because shorter tasks usually have a higher probability of completion within their deadlines. %In systems where task dropping is practiced, such as live video streaming, dropping tasks of a certain type can disrupt the system.% (such as if all resizing happened, but conversion between encoding formats was deemed too risky to be executed). 
PAMF heuristic aims at mitigating this unfairness. 

PAMF favors task types that have suffered from pruning. By relaxing the pruning thresholds for tasks of unfairly treated task types, the system can prevent bias against them. We define \emph{sufferage value} at mapping event $e$ for each task type $f$, denoted $\epsilon_{ef}$, that determines how much to decrease (\ie relax) the base pruning threshold. Note that we define 0 as no sufferage. We define \emph{fairness factor} (denoted $\vartheta$) as a constant value across all task types in a given HC system by which we change sufferage value of task types. This fairness factor denotes how quickly any task's sufferage value changes in response to missing a deadline. A high factor results in large relaxation of probabilistic requirements. Updating the sufferage value occurs upon completion of a task in the system. A successful completion of a task of type $f$ in mapping event $e$ results in lowering the sufferage value of task type $f$ by the fairness factor, \ie $\epsilon_{ef}=\epsilon_{(e-1)f}-\vartheta$, whereas for an unsuccessful task we add the fairness factor, \ie $\epsilon_{ef}=\epsilon_{(e-1)f}+\vartheta$.
We note that we limit sufferage values ($\epsilon_{ef}$) to be between 0 to 100\%. Then, the mapping heuristic determines the fair pruning threshold for a given task type $f$ at a mapping event $e$ by subtracting the sufferage value from the base pruning threshold. 

Updated pruning threshold enables PAMF create a more fair distribution of completed tasks by protecting tasks of unfairly-treated types from pruning. Once we update pruning thresholds for suffered task types, the rest of PAMF functions as PAM. 

\section{Experimental Setup}\label{sec:experiment}
\subsection{Overview}
%In this study, to reduce simulation execution times, the number of machines comprising the distributed systems is constrained, but, as the calculation of a machine's probability state can be performed at each machine, or by a control node for each subset of machines, the proposed methods are scalable to any number of machines. 
To conduct a comprehensive performance evaluation, we simulate a computing system with eight inconsistently heterogeneous machines (\ie $M=8$). %They comprise an inconsistently heterogeneous system where a given machine A can exhibit higher performance for certain task types than machine B, yet machine B may exhibit higher performance on other task types~\cite{Braun01}. 
To generate the probabilistic execution time PMFs (PET), the mean execution time results from twelve SPECint benchmarks on a set of eight machines\footnote{The 8 machines are: Dell Precision 380 3 GHz Pentium Extreme, Apple iMac
2 GHz Intel Core Duo, Apple XServe 2 GHz Intel Core Duo, IBM System X 3455 AMD
Opteron 2347, Shuttle SN25P AMD Athlon 64 FX-60, IBM System P 570 4.7 GHz,
SunFire 3800, and IBM BladeCenter HS21XM.} was determined. These mean execution times for each benchmark on each system formed the mean values for our task-machine execution times. The function describing execution time of the tasks on a machine is assumed to be a unimodal distribution; from a gamma distribution using the task-machine mean execution time, and with a shape randomly picked from the range [1:20], 500 execution times were sampled. From these times, a histogram was generated to produce a discrete probability mass function (PMF). This was repeated for each task type on each machine, and the resultant eight machine by twelve task type matrix of PMFs was stored as the PET matrix which remains constant across all of our experiments.

\subsection{Generating Workload}
Our simulation is of a finite span of time units, starting and ending in a state where the system is idle. As the system comes online, and tasks begin to accumulate in the queue, the system is not in the desired state of oversubscription. The same is true of the end of the simulation, when the last tasks are finishing, and no more are arriving to maintain the oversubscribed state. In an effort to minimize the effects of the non-oversubscribed portion of the simulation from the data, the first and last hundred (100) tasks to complete are removed from the results. Only the remaining tasks from the oversubscribed portion of the simulation are used in the analysis.

Based on other workload investigations~\cite{khemka14utility,khemka2015utility}, a gamma distribution is created with a mean arrival rate for all task types that is synthesized by dividing the total number of arriving tasks by the number of task types. 
 Except an experiment in Sub section~\ref{subsec:video}, the variance of this distribution is 10\% of the mean. Each task type's mean arrival rate is generated by dividing the number of time units by the estimated number of tasks of that type. A list of tasks with attendant types, arrivals times, and deadlines is generated by sampling each task type's distribution.
%The task types each have a mean execution time, from which the PET described above is derived. This mean execution time is used in generating the deadlines for the tasks that arrive in the system. 
For a given task $i$, the deadline is calculated as $\delta_i = arr_i + avg_i + (\beta \cdotp avg_{all})$, where \(arr_i\) is the arrival time, \(avg_i\) is the mean execution time for that task type (range from 50 to 200 ms)
, $\beta$ is a slack coefficient, and $avg_{all}$ is the mean of all task type's execution. This slack allows for the tasks to have a chance of completion in an oversubscribed system.

\subsection{Baseline Mapping Heuristics}\label{subsec:baseline}

\subsubsection{MinCompletion-MinCompletion (MM)}
This heuristic has been extensively used in the literature~\cite{he2003qos,pedemonte2016accelerating,ezzatti2013efficient, salehi2016stochastic}. In the first phase of the heuristic, the virtual queue is traversed, and for each task in that queue, the machine with the minimum expected completion time is found, and a pair is made. In the second phase, for each machine with a free slot, the provisional mapping pairs are examined to find the machine-task pair with the minimum completion time, and the assignment is made to the machine queues. The process repeats itself until all machine queues are full, or until the batch queue is exhausted.

\subsubsection{MinCompletion-Soonest Deadline (MSD)}
Phase one is as in MM. Phase two selects the tasks for each machine with the soonest deadline. In the event of a tie, the task with the minimum expected completion time is selected. As with MM, after each machine with an available queue slot receives a task from the provisional mapping in phase two, the process is repeated until either the virtual machine queues are full, or the unmapped task queue is empty.

\subsubsection{MinCompletion-MaxUrgency (MMU)}
Urgency of task $i$ on machine $j$ is defined as $U_{ij}={1}/{(\delta_i-\mathbf{E}(C_{ij}))}$, where $\mathbf{E}(C_{ij})$ is the expected completion time of task $i$ on machine $j$. 

Phase one of MMU is the same as MM. Using the urgency equation, phase two selects the task-machine pair that has the greatest urgency, and adds that mapping to the virtual queue. The process is repeated until either the batch queue is empty, or until the virtual machine queues are full.

\subsubsection{Max Ontime Completions (MOC)}
The MOC heuristic was developed in~\cite{salehi2016stochastic}. It uses the PET matrix to calculate robustness of task-machine mappings. The first mapping phase finds, for each task, the machine offering the highest robustness value. The culling phase clears the virtual queue of any tasks that fail to meet a pre-defined ($30\%$) robustness threshold. The last phase finds the three virtual mappings with the highest robustness and permutes them to find the task-machine pair that maximizes the overall robustness and maps it to that machine's virtual queue. %PAM, in contrast, chooses the pair with the minimum completion time. 
The process repeats until either all tasks in the batch queue are mapped or dropped, or until the virtual machine queues are full. %This contrasts with PAM, which defers tasks which do not meet the robustness threshold used by the system to drop tasks.

\section{Performance Evaluation}\label{sec:evltn}
\subsection{Overview}
A series of simulations were run using the Louisiana Optical Network Infrastructure (LONI) Queen Bee 2 HPC system~\cite{LONI}. For each set of tests, for each examined parameter, 30 workload trials were performed using different task arrival times built from the same arrival rate and pattern, and the mean and 95\% confidence interval of the results is reported. The arrival rates are listed in terms of number of tasks per time unit.

Each experiment is a set of 30 workload trials, consisting of 800 tasks per trial. Each of the experiments investigates extreme levels of oversubscription  where few tasks complete successfully using baseline heuristics. Each machine in the HC system has a machine-queue size of six, counting the executing task and the dropping toggle is one task. For each of the experiments, unless otherwise noted, the performance metric (and the vertical axis) is the percentage of tasks completed before their deadline (\ie overall robustness).

\subsection{Dynamic engagement of probabilistic task dropping}\label{subsec:toggle}

\begin{figure} [ht]
\vspace{-10pt}
  \centering
  \includegraphics[width=0.33\textwidth]{\paperfolder/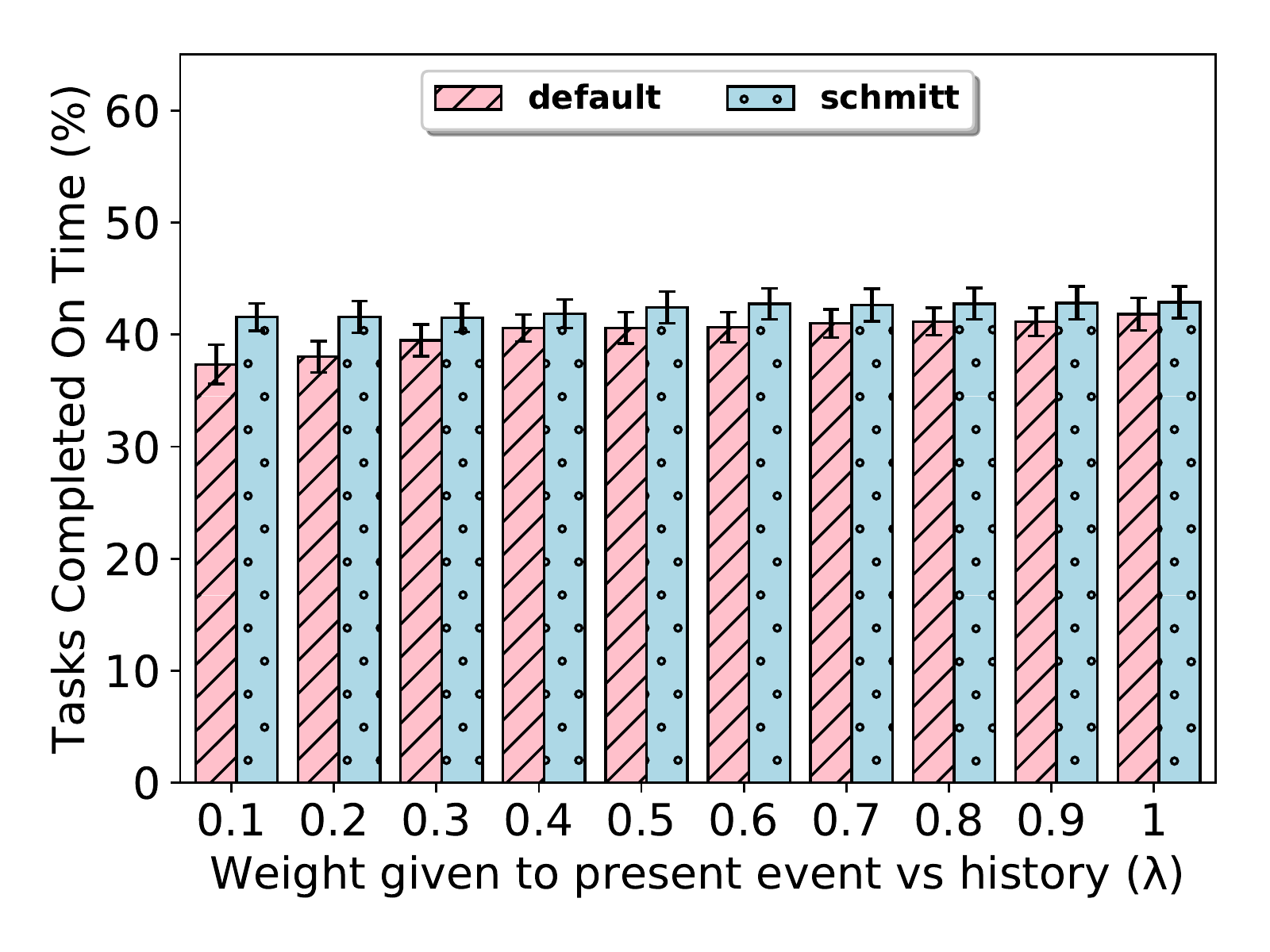}
  \caption{\small{Impact of historical oversubscription observations and Schmitt Trigger on determining oversubscription level of HC system.} \label{fig:schmitt} }
  \vspace{-10pt}
\end{figure}

In this experiment, our aim is to appropriately configure oversubscription level (see Equation~\ref{eq:oversub}) by determining the weight that should be assigned to the number of deadlines missed in the recent mapping event versus the previous values of the oversubscription level. We also evaluate the impact of using Schmitt Trigger as opposed to using a single threshold for dynamic engagement of task dropping. This experiment was conducted under 34k tasks arriving to the system. 
%MOHSEN COMMENT FOR A LATER TIME: WHY NOT 34K ARRIVAL RATE?

Figure~\ref{fig:schmitt} shows that by assigning a higher weight to the number of dropped tasks in the most recent mapping event, the overall robustness of the system is increased from 67.7\% to 71.9\%. This is due in part to the steady nature of task-arrival our workload trials with only few sudden spikes. Also, we can observe that applying Schmitt Trigger results in a higher robustness. Specifically, we observe that $\lambda=0.9$ provides a statistically and practically higher robustness, hence, is appropriate for identifying oversubscription level.

We can conclude that under high oversubscription levels, the best results come from taking immediate action when tasks miss their deadlines, and then a steady application of probabilistic task dropping until the situation is decidedly controlled (\ie reaching the lower bound of Schmitt Trigger).

\begin{comment}
\subsection{Impact of per-task dynamic dropping threshold}
\begin{figure} [ht]
\vspace{-10pt}
  \centering
  \includegraphics[width=0.35\textwidth]{\paperfolder/Figures/dynamic.pdf}
  \caption{\small{Impact of dynamic per-task dropping thresholds. Each bar shows a different $\rho$ value (the weight of the dynamic dropping threshold). The horizontal axis is the level of oversubscription in form of number of tasks.}} \label{fig:dynamic}
  \vspace{-10pt}
\end{figure}

Figure \ref{fig:dynamic} shows that the per-time thresholds result in no statistical difference in robustness. The machines and tasks in this work come from benchmarks, which coupled with the arrival rates and patterns used, result in PMFs for task-machine mappings with performance characteristics that hide the potential effectiveness of such a system. %To more deeply analyze the potential benefits of a per-task dropping threshold, further research needs to be done using carefully crafted tasks/machines/arrival patterns.
\end{comment}

\subsection{Impact of deferring and dropping thresholds}\label{subsec:thresh}
\begin{figure} [ht]
\vspace{-10pt}
  \centering
  \includegraphics[width=0.33\textwidth]{\paperfolder/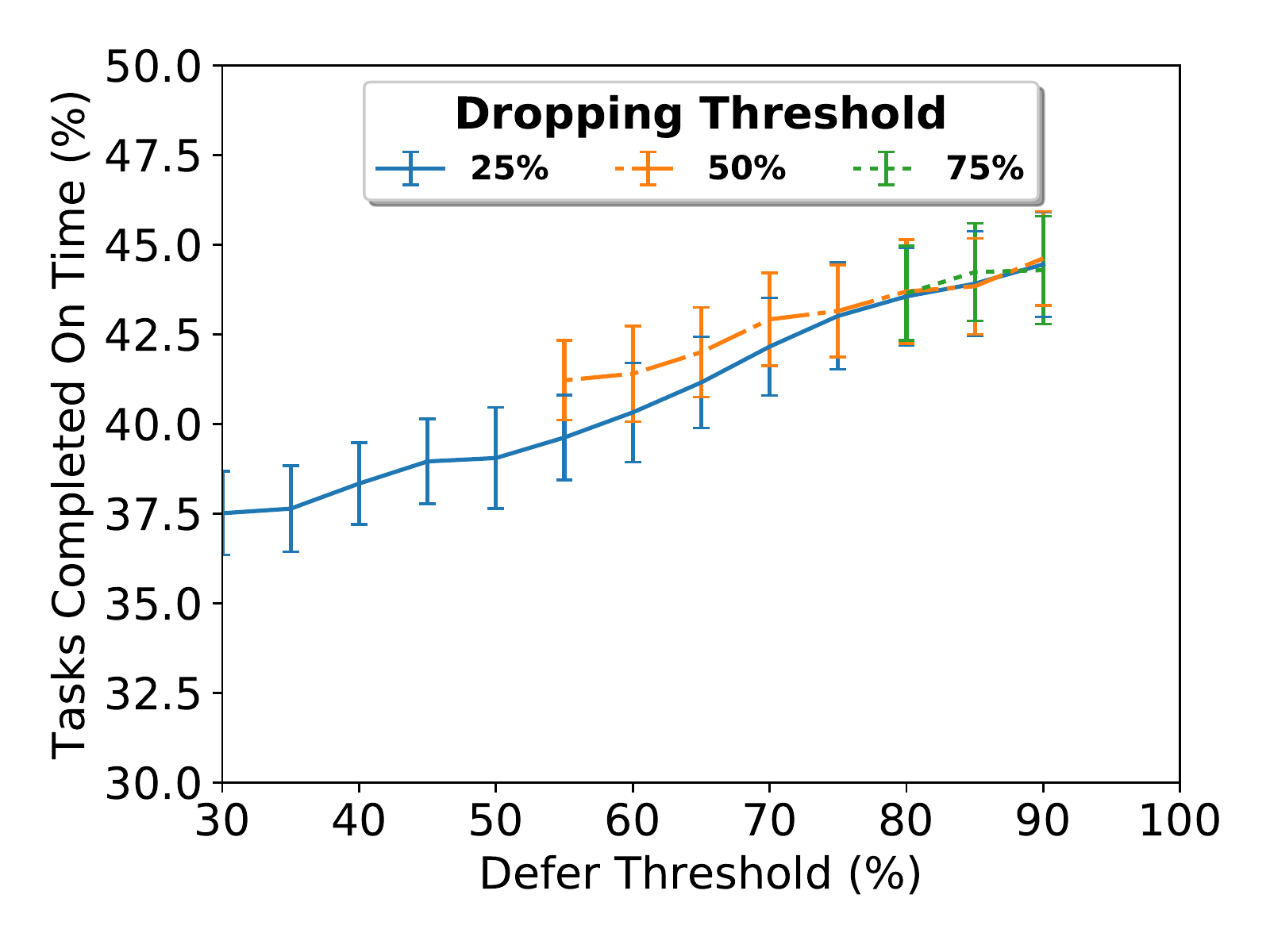}
  \caption{\small{Impact of deferring and dropping thresholds. Dropping threshold is denoted by line type and color. }} \label{fig:decouple} 
  \vspace{-10pt}
\end{figure}

%As we discussed in , deferring threshold must be higher than the dropping threshold. 
The goal of this experiment is to identify the appropriate gap that should be considered between the deferring and dropping thresholds (see Section~\ref{subsec:dropdefer}). To find the appropriate deferring threshold, we add a gap value to the dropping threshold (\eg a dropping threshold of 50\% would require at least 55\% robustness to map a task to a machine). To test this, three dropping thresholds (25\%, 50\%, and 75\%) are examined in an experiment increasing the gap on each by 5\% until the deferring threshold reaches 90\%. The results, shown in Figure~\ref{fig:decouple}, are generated from an a workload with 34k tasks. %

Figure~\ref{fig:decouple} validates our hypothesis by showing that using a higher deferring threshold leads to higher system robustness. 
We observe that if the deferring threshold is chosen high enough, the impact of dropping threshold on the system robustness diminishes. That is, with a high dropping threshold (\eg 75\%) we obtain the same system robustness with a low dropping threshold (\eg 25\%). However, we note that higher dropping threshold can influence the incurred cost of using an HC system, because they prevent wasting time processing unlikely-to-succeed tasks that have been mapped to the system. In the rest of experiments, dropping threshold 50\% and deferring threshold 90\% is used.

\subsection{Evaluating the impact of fairness factor}
\begin{figure} [ht]
\vspace{-10pt}
  \centering
  \includegraphics[width=0.33\textwidth]{\paperfolder/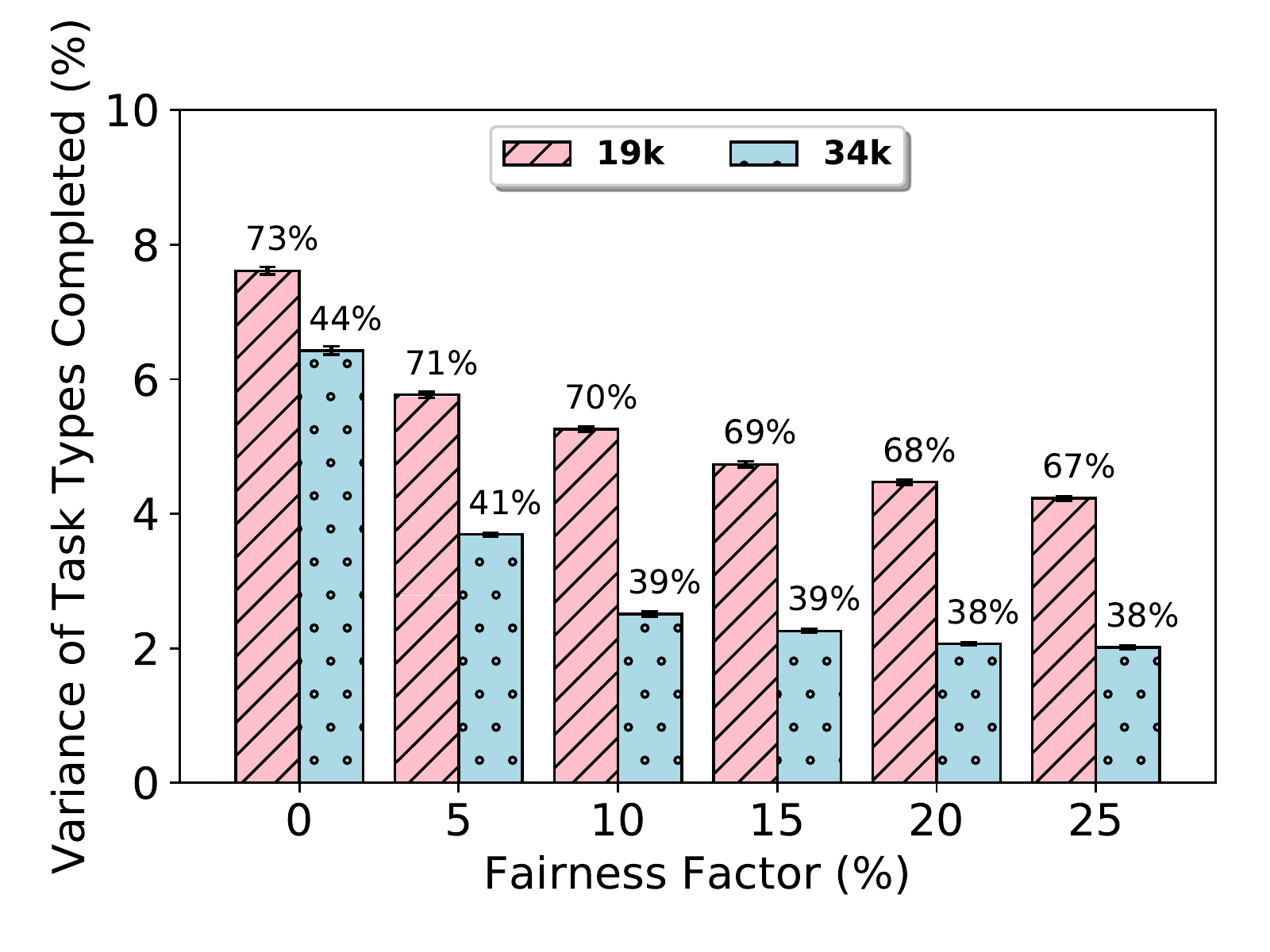}
  %\caption{Percentage of tasks meeting their deadlines (vertical axis) examining the impact of different levels of task-fairness adjustment. Dropping toggle is set to 1 missed task. Pruning threshold is set to 75\%. Horizontal axis shows percentage added to the tracked per-task pruning threshold when tasks of that type fail to complete on time. Each bar is labeled with the system robustness under that fairness implementation. \label{fig:fairness} }
  \caption{\small{Evaluating fairness and robustness. Horizontal axis shows fairness factor modifier to the sufferage value. Vertical axis is the variance of completed task types. Values above bars show robustness.} }
  \label{fig:fairness} 
  \vspace{-10pt}
\end{figure}

Our aim is to study if PAMF heuristic (see Section~\ref{sec:solution}) alleviates unfairness. We test the system using a fairness factor ranging from 0\% (\ie no fairness) to 25\%. Recall that this fairness factor is the amount by which we modify the sufferage value for each task type. The sufferage value for a given task type at a given mapping event is subtracted from the required threshold, in an effort to promote fairness in completions amongst task types. For each fairness factor, we report: (A) The variance in percentage of each task type completing on time. The objective is to minimize the variance among these. (B) The overall robustness of the system, to understand the robustness we have to compromise to attain fairness. Robustness value is noted above each bar in Figure~\ref{fig:fairness}. We tested oversubscription level of 19k and 34k tasks.

Figure \ref{fig:fairness} shows that significant improvement in fairness can be attained at the cost of compromising robustness. In particular, we observe that using 5\% fairness factor results in remarkable reduction  in variance of completed tasks that implies increasing fairness. For instance, for 34k tasks, the variance drops from 6\% to 3.5\%, at a cost of $\simeq$10\% reduction in robustness (from 44.2\% to 40.5\%). This compromise in robustness is because deferring fewer tasks in an attempt to improve fairness results in fewer tasks successfully completed overall. Further increasing fairness factor results in insignificant changes on fairness and robustness, therefore, we configure PAMF with 5\% fairness factor in the experiments. 

\subsection{Evaluating robustness of pruning}
In this experiment, we compare the overall robustness offered by PAM and PAMF against that of the baseline heuristics described in section~\ref{sec:experiment}. We conducted this evaluation under various oversubscription levels, however, due to space limit and presentation clarity we only show oversubscription levels with 19k and 34k tasks. We note that the same pattern is observed with other oversubscription levels evaluated.

\begin{figure} [ht]
\vspace{-10pt}
  \centering
  \includegraphics[width=0.33\textwidth]{\paperfolder/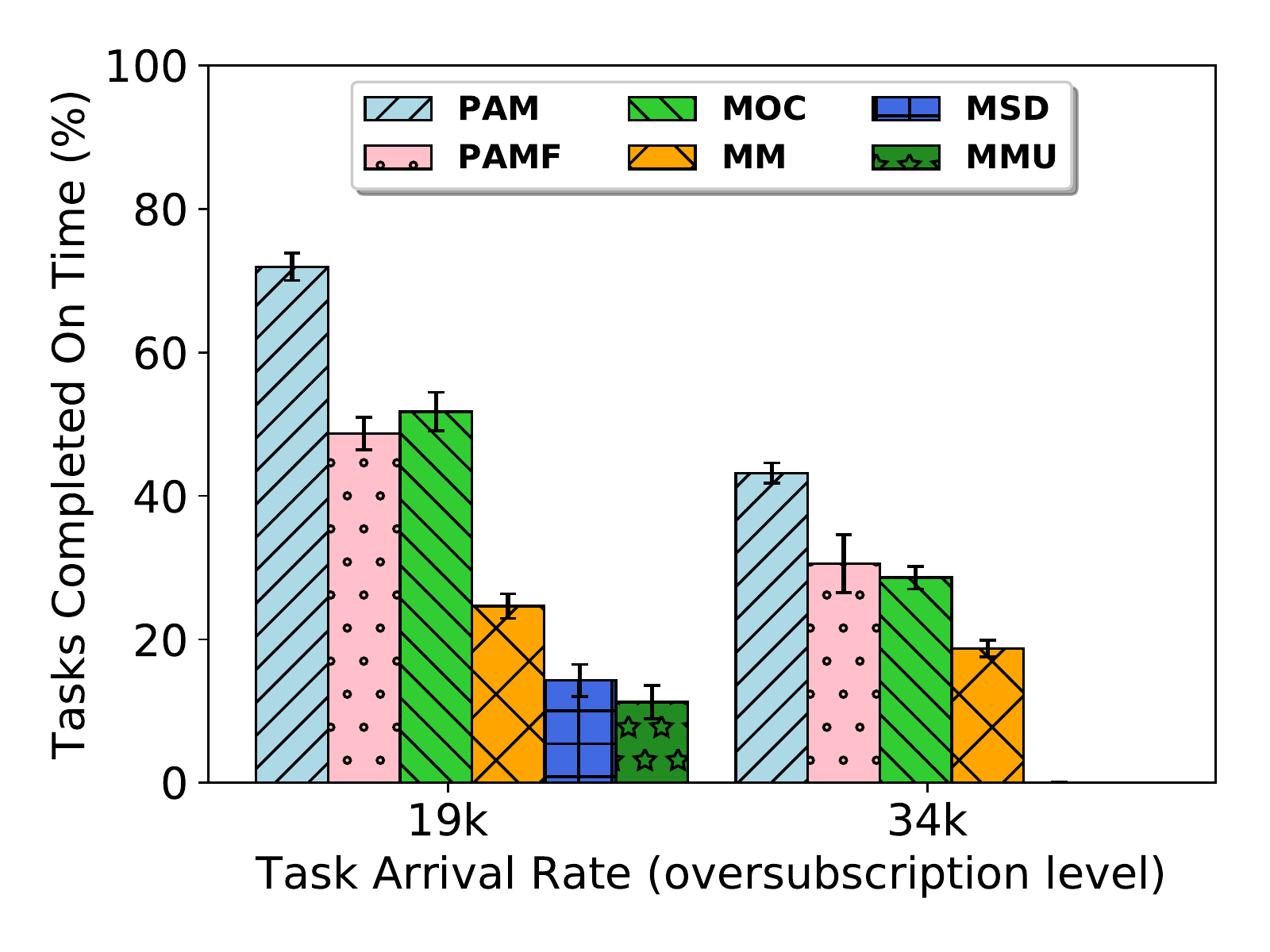}
  \caption{\small{Comparison of PAM and PAMF against other heuristics. Horizontal axis shows oversubscription in form of number of tasks.}}
   \label{fig:compare} 
  \vspace{-10pt}
\end{figure}

In Figure~\ref{fig:compare}, we observe that PAM results in a substantial increase in system robustness versus other heuristics, at nearly 70\% and PAMF results in nearly 50\% robustness, trading percentage of tasks completed for types of tasks completed. MOC, another robustness-based heuristic, is the closest in robustness to PAM, rivaling PAMF, at nearly 50\%. The inability to probabilistically drop tasks leads to wasted processing and delayed task mapping, thereby lowering robustness. With robustness of $\simeq$25\%, the performance of MinMin lags far behind, as it allocates tasks to machines no matter how unlikely they are to succeed. The robustness offered by both MSD and MMU suffers in comparison because these heuristics, instead of maximizing performance of the most-likely tasks, prioritize tasks whose deadlines or urgency is closest (\ie least likely to succeed tasks). With an oversubscription of 34k tasks, MSD and MMU only map tasks that fail to meet their deadline.%, with no ability to recover from an initial mistaken mapping.

\subsection{Cost benefit of probabilistic pruning}
%While the focus of this paper has been maximizing robustness in an HC system, there are other metrics of success to consider; one of these is cost. Time spent computing tasks that fail to successfully complete is a waste of computing resources which for certain scenarios, such as cloud computing, have associated costs. Dropping threshold is 50\% and deferring threshold is 90\%. A $\lambda$ of 1 is used.

\begin{figure} [ht]
\vspace{-10pt}
  \centering
  \includegraphics[width=0.33\textwidth]{\paperfolder/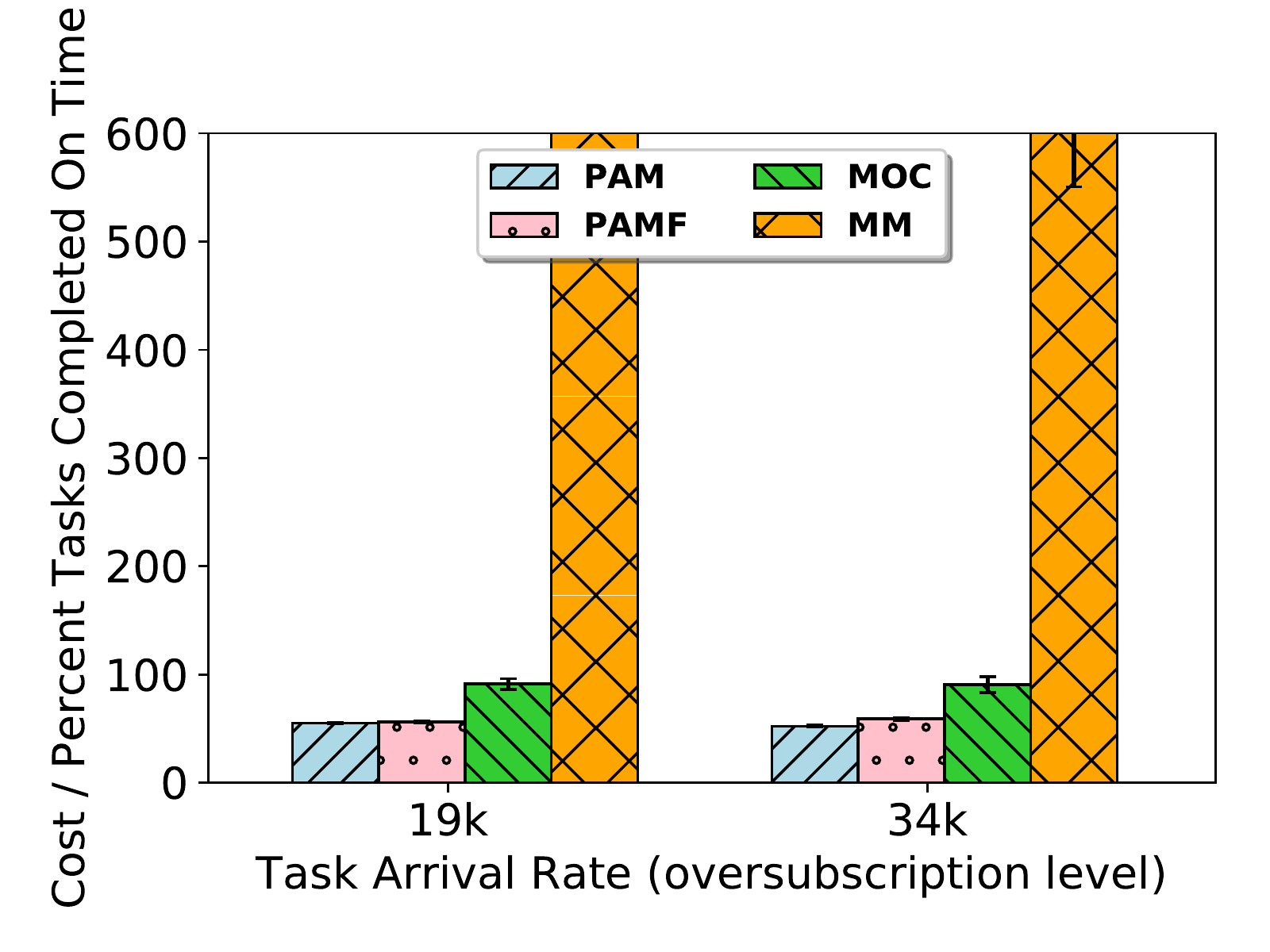}
  \caption{\small{Impact of probabilistic pruning on incurred costs of using resources. Horizontal axis shows the oversubscription level.} }\label{fig:cost} 
  \vspace{-10pt}
\end{figure}

To investigate the incurred cost of using resources, pricing from Amazon cloud VMs~\cite{aws} has been mapped to the machines in the simulation. Each machine's usage time is tracked. The price incurred to process the tasks is divided by the percentage of on-time tasks completed to provide a normalized view of the incurred costs in the system.

Figure \ref{fig:cost} shows that in an oversubscribed system both PAM and PAMF incur a significantly ($\simeq$40\%) lower cost per completed task than MOC and other heuristics. %When the arrival rate is at 9k, PAM performs a little worse, as in an effort to ensure the highest success rate, the best machines are always chosen (\ie often the most expensive machines), resulting in the same number of tasks completed for a higher cost. However, as the arrival rate increases, the relative cost of PAM decreases. 
At extreme levels of oversubscription, the difference between heuristics such as MMU and MSD and PAM become unchartable, as MSD and MMU both prioritize tasks least likely to succeed, whereas PAM prioritizes those most likely to succeed. 
%In the highest tested levels of oversubscription, MMU performed so poorly that zero tasks completed in any trial, resulting in a cost of zero. 
While previous tests have shown PAM outperforms other heuristics in terms of robustness in the face of oversubscription, these results show that in most levels of oversubscription, the benefits are realized in dollar cost as well, due to not processing tasks needlessly.

\subsection{Evaluating video transcoding workload traces}
  \label{subsec:video} 

\begin{figure} [ht]
\vspace{-12pt}
  \centering
  \includegraphics[width=0.32\textwidth]{\paperfolder/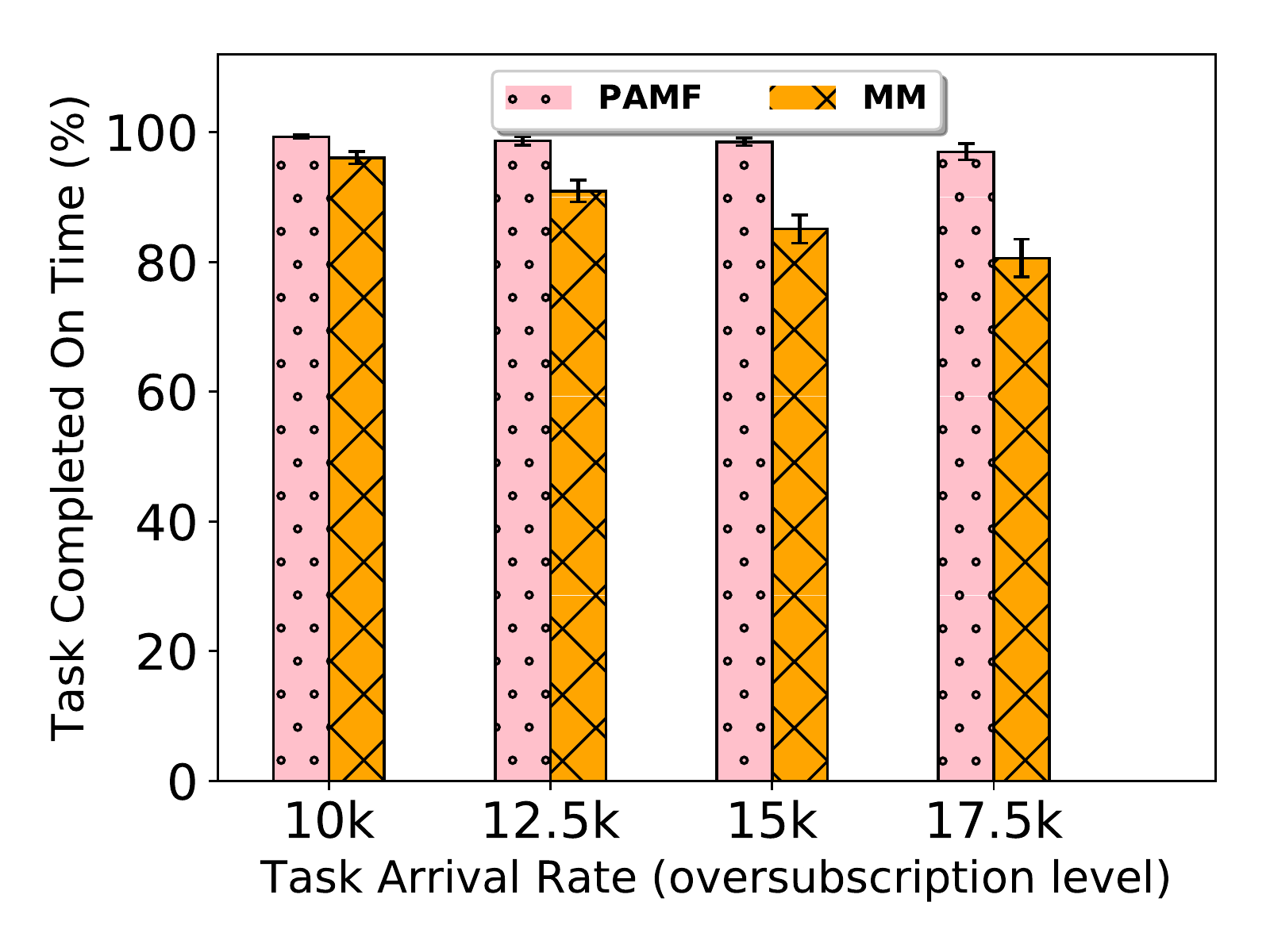}
  %\caption{Percentage of tasks meeting their deadlines (vertical axis) examining the impact of different levels of task-fairness adjustment. Dropping toggle is set to 1 missed task. Pruning threshold is set to 75\%. Horizontal axis shows percentage added to the tracked per-task pruning threshold when tasks of that type fail to complete on time. Each bar is labeled with the system robustness under that fairness implementation. \label{fig:fairness} }
    \vspace{-5pt}
  \caption{\small{Comparison of PAMF against MinMin using video transcoding workload. The horizontal axis shows the oversubscription level (number of tasks).} }
  \label{fig:video} 
  \vspace{-10pt}
\end{figure}

To evaluate PAMF under a real-world setting, we compare it against MinMin on video transcoding workload traces under different oversubscription levels (horizontal axis in Figure~\ref{fig:video}
). The PET matrix captured from running four video transcoding types on 660 video files (available in \url{https://goo.gl/TE5iJ5}) on four heterogeneous Amazon EC2 VMs, namely CPU-Optimized, Memory-Optimized, General Purpose, and GPU. The experimental result confirms our earlier observation and show that PAMF outperforms MinMin specifically as the level of oversubscription increases.

\section{Conclusion and Future Works}\label{sec:conclsn}
The goal of this research was to improve robustness of HC systems via pruning tasks with low probability of success. We designed a pruning mechanism as part of resource allocation system in the system. For pruning, we determined probability values used by mapping heuristics to either map or defer a task. We concluded that: \textbf{(A)} when the system is not oversubscribed, tasks with low chance of success should be deferred (\ie wait for more favorable mapping in the next mapping); \textbf{(B)} When the system is sufficiently oversubscribed, the unlikely-to-succeed tasks must be dropped to alleviate the oversubscription and increase the probability of other tasks succeed; \textbf{(C)} The system benefits from setting higher deferring threshold than dropping threshold. We developed a mapping heuristic, PAM, based on the probabilistic task pruning and showed that it can improve system robustness by on average by $\simeq$25\%. We upgraded PAM to accommodate fairness by compromising around four percentage points robustness. Evaluation results revealed that pruning mechanism (and PAM) does not only improve system robustness but also reduces the cost of using cloud-based HC systems by $\simeq$40\%.

The idea of pruning developed in this research is generic and can be plugged to other systems. We plan to extend the probabilistic approach for tasks preemption and its impact on the convolution process. Another future work is to approximately compute tasks, in addition to pruning or dropping them from the system. Finally, as HC systems have various QoS concerns, domain-specific fairness models should be explored.

\section*{Acknowledgments}
We thank reviewers of the paper.
Portions of this research were conducted with high performance computational resources provided by the Louisiana Optical Network Infrastructure~\cite{LONI}.
This research was supported by the Louisiana Board of Regents under grant number LEQSF(2016-19)-RD-A-25. 

 \linespread{0.97}
\bibliographystyle{IEEEtran}
\balance
\bibliography{references}

\end{document}